\documentclass[a4paper,10pt]{article}
\usepackage{graphicx}
\usepackage[left=2cm,right=2cm,
top=2cm,bottom=2cm,bindingoffset=0cm]{geometry}
\usepackage{amssymb}
\usepackage{amsmath}
\usepackage[classfont=bold,langfont=roman,funcfont=roman]{complexity}
\usepackage{cite}
\usepackage{xcolor}
\def\be{\begin{eqnarray}}
\def\ee{\end{eqnarray}}

\title{Regularization of $\delta'$ potential in general case of deformed space with minimal length }
\author{M. I. Samar and V. M. Tkachuk\\ Professor Ivan Vakarchuk Department for Theoretical Physics, \\ Ivan Franko National University of Lviv,\\ 12 Drahomanov St, Lviv,
UA-79005, Ukraine}
\begin{document}

\maketitle

\begin{abstract}
In general case of deformed Heisenberg algebra leading to the minimal length we present a definition of  the  $\delta'(x)$ potential as a linear kernel of potential energy operator in momentum represenation.
We find exactly the energy level and  corresponding eigenfunction  for $\delta'(x)$ and $\delta(x)-\delta'(x)$  potentials  in deformed space with arbitrary function of deformation. The energy spectrum for different partial cases of deformation function is analysed. 



\end{abstract}

\section{Introduction}

String theory and quantum gravity  suggest the existence of minimal length as a finite lower bound to the possible resolution of length \cite{GrossMende,Maggiore,Witten}.
 Kempf et al. showed that minimal length can be achieved by modifying usual canonical commutation relations \cite{Kempf1994,KempfManganoMann,HinrichsenKempf,Kempf1997}.
One of the simplest deformed Heisenberg algebra in one-dimensional case is the one proposed by Kempf  \cite{Kempf1994}
\be \label{Kempf}
[\hat{X},\hat{P}]=i\hbar (1+\beta \hat{P}^2),
\ee
leading to minimal length $\hbar\sqrt{\beta}$.

More general deformed algebra has the form
\be \label{general_deformation}
[\hat{X},\hat{P}]=i\hbar f({\hat{P}}),
\ee
where $f$ is called function of deformation, being strictly positive ($f >0$), even function. Algebra (\ref{general_deformation}) admits the following representation
\be\label{psevdo-position}
&&{\hat{X}}=\hat{x}=i\hbar\frac{d}{dp},\\ \nonumber
&&{\hat{P}=g({p})}.
\ee
Function $g({p})$  is an odd function  satisfying $\frac{dg(p)}{dp}=f(g(p))$.
It is defined on finite domain $ p \in [-b,b]$, with $b=g^{-1}(a)$.
Here $a$ denotes the limit of momentum $P \in [-a,a]$. Note  that finiteness of $b$ provides the existence of minimal uncertainty in position \cite{Nowicki}.

The  study of the effect of the minimal length on systems with singular potentials or point interactions is of particular interest, since such systems are expected to have a  nontrivial sensitivity to minimal length. The impact of the minimum length  has been studied in the context of the following   problems with singularity in potential energy:   hydrogen atom \cite{Brau,Benczik,StetskoTkachuk,Stetsko2006,Stetsko2008, SamarTkachuk, Samar,Quesne:2010}, gravitational quantum well \cite{Brau2006, Nozari2010, Pedram2011}, a particle in delta potential and double delta potential\cite{Samar1, Ferkous}, one-dimensional Coulomb-like problem \cite{Samar1,Fityo,Samar2}, particle in the singular inverse square potential \cite{Bouaziz2007,Bouaziz2008, Bouaziz2017,Samar2020}, two-body problems with delta and Coulomb-like interactions \cite{Samar2017}. 

 In undeformed quantum mechanics the interest in studies of point interactions is twofold. 
The first reason is that point interaction is a  good model of a very localized interaction connected with  different structures like quantum waveguides \cite{Albeverio2007,Cacciapuoti}, spectral filters \cite{Turek1,Turek2}, or infinitesimally thin sheets \cite{Zolotaryuk1,Zolotaryuk2}.
Another reason is that it is often possible for such systems to obtain the solution exactly.

The problem of the correct interpretation of the hamiltonian with $\delta'$-function in potential energy   \be \label{Ham} \hat{H}=-\frac{\hbar^2}{2m}\frac{d^2}{dx^2}-\kappa\delta'(x)\ee
has been considered in literature since the 80s of last century\cite{Albeverio,Seba1,Seba2}. 
The $\delta'$-potential  is  very  sensitive to a way of its regularization.  From a physical point of view, this means that there is no unique one-dimensional model of the delta prime interaction described by the hamiltonian (\ref{Ham}). 
In order to avoid any confusion it should be emphasized that there can be distinguished a few different approches  corresponding to $\delta'$-potential definition.

 By the first time $\delta'$-interaction was considered in \cite{Albeverio}.  
  The hamiltonian $H$ was defined as the one-parameter family of self-adjoint extensions of an operator
 \be H_{\beta}=-\frac{\hbar^2}{2m}\frac{d^2}{dx^2}\ee acting on the domain of wavefunction  with derivative to be continuous, while the wavefunction has a jump proportional to its derivative at $x=0$
 \be\label{bound_cond}\psi'(-0)=\psi'(+0), \  \ \  \psi(+0)-\psi(-0)=\beta \psi'(0),  \ee
 with $\beta$ depending on $\kappa$, defined in (\ref{Ham}). 
It was shown in \cite{Seba2} that this selfadjoint extensions correspond to the heuristic operator 
\be \label{Seba1}
 H_{\beta}=-\frac{\hbar^2}{2m}\frac{d^2}{dx^2} +\beta |\delta'><\delta'|, 
\ee
with renormalized coupling. Here $|\delta'><\delta'|$ denotes the following operator
\be (|\delta'><\delta'| \psi)(x) =\delta'(x)\int\delta'(y)\psi(y)dy.\ee
  Operator  (\ref{Seba1}) is not very good to describe the $\delta'$-potential, thus.
 
In \cite{Kurasov} it was proposed to   define  self-adjoint operator (\ref{Ham})  using distribution theory for discontinuous functions and derive the following boundary conditions at the point where the interaction occurs 
 \be \psi(+0)-\psi(-0)=\frac{\kappa}{2}\left(\psi(+0)+\psi(-0)\right) \\ \psi'(+0)-\psi'(-0)=-\frac{\kappa}{2}\left(\psi'(+0)+\psi'(-0)\right).  \ee
 
 In the same paper \cite{Seba2} the alternative definition of the problem was proposed  
 \be
 H_{\beta}=-\frac{\hbar^2}{2m}\frac{d^2}{dx^2} +\frac{\kappa}{\varepsilon^\alpha}(\delta(x+\varepsilon)+\delta(x-\varepsilon)).
 \ee
 Hovewer, Seba has proved that in the limit of $\varepsilon\rightarrow 0$ the interaction disappears if $\alpha<1/2$, appears as a $\delta(x)$ potential for $\alpha=1/2$ and splits the system into two independent subsystems lying on the half-lines $(-\infty,0)$ and $(0,\infty)$ if $\alpha>1/2$.

 The one more way of defining  the Schr\"odinger operator with a potential $\delta'$ is to approximate $\delta'$ by regular potentials  and then to investigate the convergence of the corresponding family of regular Schr\"odinger operators. This approach was firstly  realized in \cite{Seba2} and studied in \cite{Manko, Golovaty}.

The aim of this paper is to show that in case of generalized uncertainty principle it is possible to introduce $\delta'(x)$ potential in some natural way for deformed space with minimal length. Our proposal leads to the regularization of the divergent integrals associated with the energy levels of the  $\delta'(x)$ potential in one-dimensional nonrelativistic quantum mechanics. 

We organize the rest of this paper as follows. In Section 2, we propose the definition of  $\delta'(x)$ potential in momentum representation in general case of deformed space with minimal length and obtain the exact relation from which the corresponding bound states energies can be extracted . In Section 3,  the Schr\"odinger equation with   $\delta(x)-\delta'(x)$ potential in context of minimal length assumption  is solved exactly. Some concluding remarks are reported in the last section.
\section{$\delta'$ potential}
In general case of deformed space with minimal length   Schr\"odinger equation in the momentum representation can be written as
\be \label{1}
\frac{g^2(p)}{2m}\phi(p)+\int_{-b}^{b}U(p-p')\phi(p')dp'=E\phi(p)
\ee
with $U(p-p')$ being the kernel of the potential energy operator.
In undeformed space this kernel can by obtained by
\be \label{connection}
U(p-p')=\frac{1}{2\pi\hbar}\int_{-\infty}^{\infty}V(x)\exp\left(-\frac{i}{\hbar}(p-p')x\right)dx.
\ee
In case of the delta prime interaction $V(x)=\kappa\delta'(x)$
the kernel of potential energy is
\be\label{kernel_dp}
U(p-p')=\frac{i{\kappa}}{2\pi\hbar^2}(p-p').
\ee 
We assume that in deformed space with minimal length $U(p-p')$ is still expessed by formula (\ref{kernel_dp}) and write the Schr\"odinger equation for the delta prime potential in general case of deformed space with minimal length as 
\be\label{Schr_dp}
(g(p)^2+q^2)\phi(p)+\frac{i{\kappa}m}{\pi\hbar^2}\int_{-b}^{b}(p-p')\phi(p')dp'=0.
\ee
The solution of (\ref{Schr_dp}) can be proposed   in the form
\be\label{f_dp}
\psi(p)=\frac{Ap+B}{g^2(p)+q^2}.
\ee
Substituting (\ref{f_dp}) into  (\ref{Schr_dp}) we obtain the following formulas
\be
A+\frac{i{\kappa}m}{\pi\hbar^2}B\int_{-b}^{b}\frac{dp}{g^2(p)+q^2}=0,\\
B-\frac{i{\kappa}m}{\pi\hbar^2}A\int_{-b}^{b}\frac{p^2dp}{g^2(p)+q^2}=0,
\ee
which yield the equation for energy spectrum
\be\label{delta_prime_energy}
1=\alpha I_1(\varepsilon)I_2(\varepsilon)
\ee
where $\alpha=\frac{\kappa^2 m^2}{\pi^2\hbar^4}$ and 
\be
I_1(\varepsilon)=b\int_{-b}^{b}\frac{dp}{g^2(p)+q^2}=\int_{-1}^1\frac{dy}{k^2(y)+\varepsilon^2}\\
I_2(\varepsilon)=\frac{1}{b}\int_{-b}^{b}\frac{p^2dp}{g^2(p)+q^2}=\int_{-1}^1\frac{y^2dy}{k^2(y)+\varepsilon^2}
\ee
with $y=\frac{p}{b}\in[-1,1]$, $g(p)=bk(y)$ and $\varepsilon=\frac{q}{b}$.

It is important to note that in undeformed limit $b\rightarrow\infty$ integral $I_1(\varepsilon)$ diverges and Schr\"odinger equation (\ref{1}) has no solution. In case of finite $b$ integrals in (\ref{delta_prime_energy}) are convergent and the problem of delta prime potential is regularized in deformed space with minimal length. Both integrals $I_1(\varepsilon)$ and $I_2(\varepsilon)$ are positive and decreasing functions of $\varepsilon>0$. This means that equation (\ref{delta_prime_energy}) has only one solution since $\alpha$ can take only positive values.

Let us find the energy spectra of considerable problem for some special examples of deformation function.

\textbf{Example 1.} 

In the simplest deformed commutation relation leading to minimal length
\be \label{cutoff}
f(P)=1, \ \ P\in[-b,b], \ \  g(p)=p.
\ee
Integrals $I_1(\varepsilon)$ and $I_2(\varepsilon)$ are
\be
&&I_1(\varepsilon)=\frac{2\arctan\left(\frac{1}{\varepsilon}\right)}{\varepsilon}, \\
&&I_2(\varepsilon)=2-2\varepsilon\arctan\left(\frac{1}{\varepsilon}\right).
\ee
Equation for energy spectrum reads
\be
1=4\alpha \arctan\left(\frac{1}{\varepsilon}\right)\left(\frac{1}{\varepsilon}-\arctan\left(\frac{1}{\varepsilon}\right)\right).
\ee
\begin{figure}[h!]
	\centering
	\includegraphics[width=14 cm]{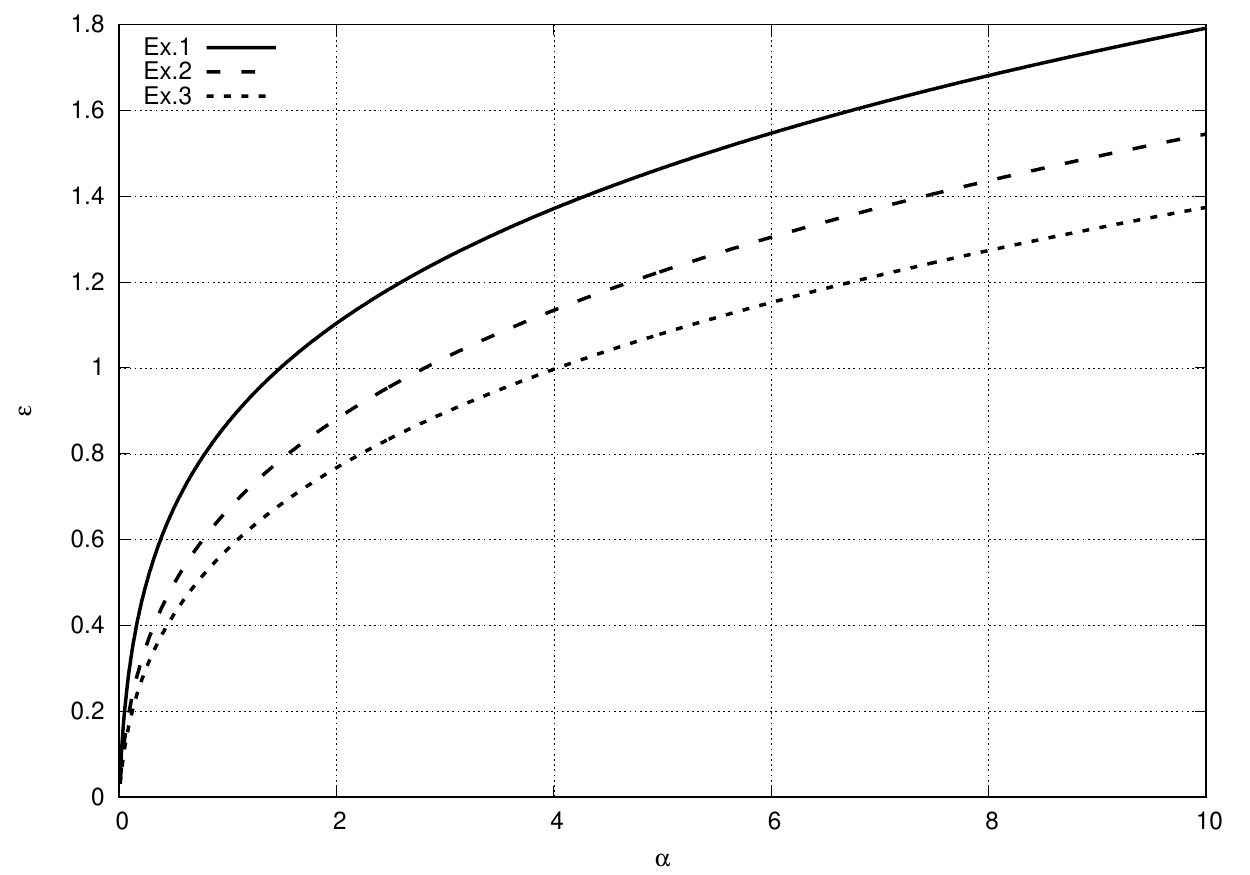}
	\vspace{-10 pt}
	\caption{\footnotesize{Energy level of delta prime problem in deformed space with minimal length dependent on the coupling constant $\alpha$}.}
	\label{fig1}
\end{figure}
\textbf{Example 2.} 
The next example of  deformation function is the following
\be \label {ex2}f(P)=(1+\beta P^2)^{3/2},\ \ \ a=\infty,\\
g(p)=\frac{p}{\sqrt{1-\beta p^2}},\ \ \ \ \ b=\frac{1}{\sqrt{\beta}}.\ee
The needed integrals can be calculated explicitly
\be
&&I_1(\varepsilon)=-\frac{2}{1-\varepsilon^2}+\frac{2\arctan\left(\frac{\sqrt{1-\varepsilon^2}}{\varepsilon}\right)}{\varepsilon(1-\varepsilon^2)^{3\over2}}, \\
&&I_2(\varepsilon)=\frac{2(\varepsilon^2+2)}{3(1-\varepsilon^2)^2}-\frac{2\varepsilon\arctan\left(\frac{\sqrt{1-\varepsilon^2}}{\varepsilon}\right)}{(1-\varepsilon^2)^{5\over2}}.
\ee

\textbf{Example 3.} In case of Kempf's deformation
\be \label {ex2}f(P)=(1+\beta P^2),\ \ \ a=\infty,\\
g(p)=\frac{1}{\sqrt{\beta}}\tan(\sqrt{\beta}p),\ \ \ \ \ b=\frac{\pi}{2\sqrt{\beta}}\ee
integrals $I_1(\varepsilon)$ and $I_2(\varepsilon)$ can also be calculated
\be
&&I_1(\varepsilon)=\frac{2\pi}{\varepsilon(\pi\varepsilon+2)}, \\
&&I_2(\varepsilon)=\frac{2}{3}\frac{\varepsilon\pi^3 - 12\polylog\left(2,\frac{\pi\varepsilon-2}{\pi\varepsilon+2}\right)+12\polylog\left(2,\frac{\pi\varepsilon+2}{\pi\varepsilon-2}\right)}{\pi\varepsilon(\pi^2\varepsilon^2-4)}.
\ee
The comparison of the dependencies of  the energy $\varepsilon$ on coupling constant $\alpha$ is presented on Fig. \ref{fig1}.

\section{$\delta-\delta'$ potential}
In this section let us consider  more general potential in the form 
\be
V(x)=-\lambda \delta(x)+\kappa\delta'(x)
\ee
The kernel of the potential energy operator is the following
\be
U(p-p')=-\frac{\lambda}{2\pi\hbar}+\frac{i{\kappa}}{2\pi\hbar^2}(p-p').
\ee
\begin{figure}[h!]
	\centering
	\includegraphics[width=8.4cm]{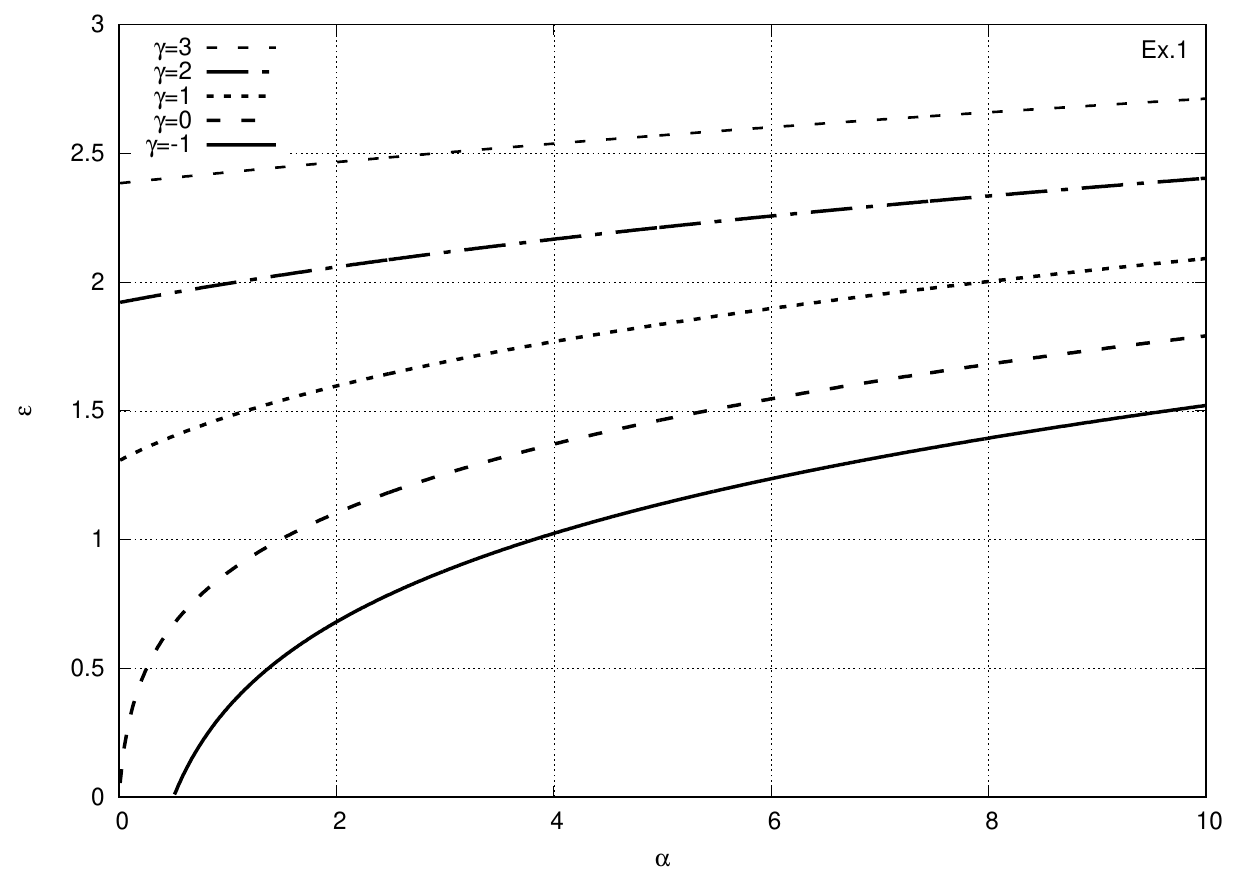}
	\includegraphics[width=8.4 cm]{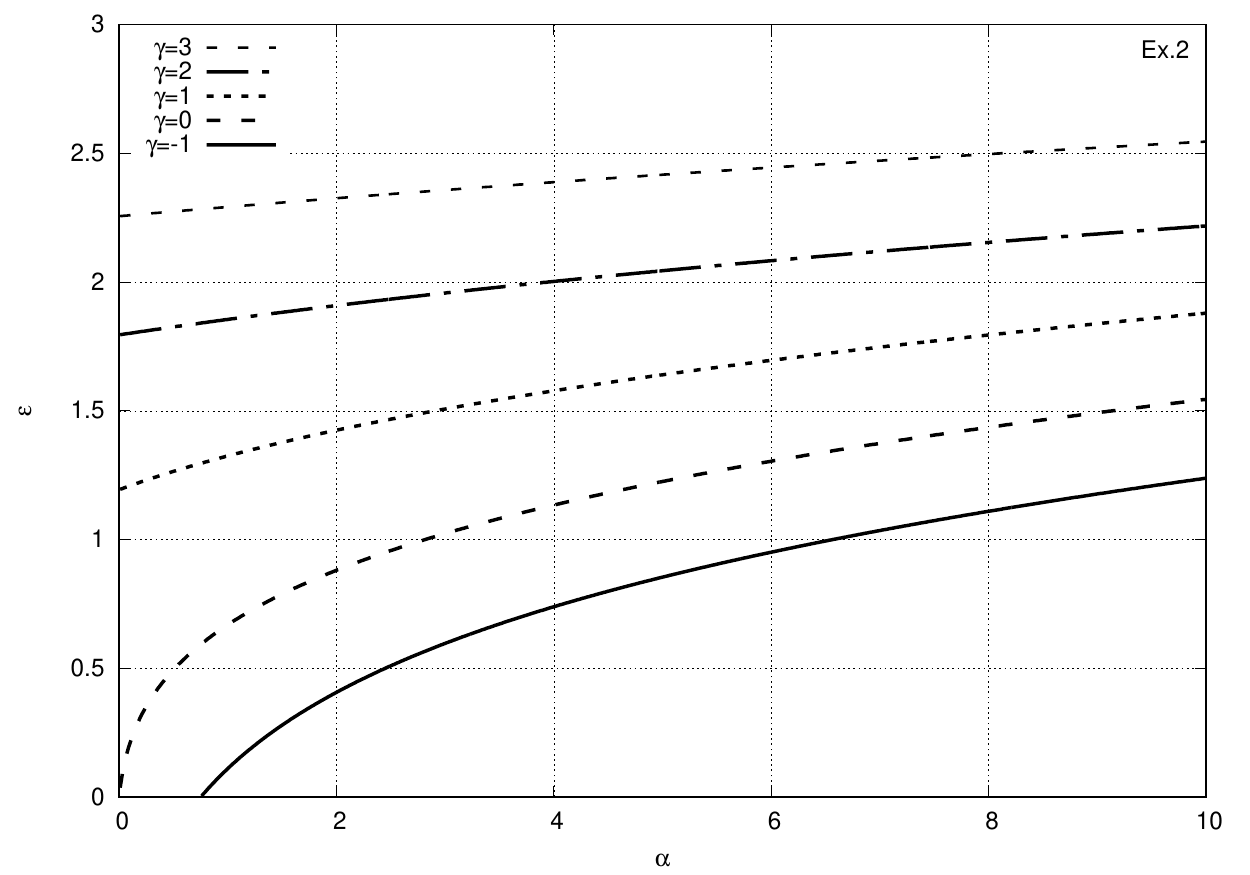}
	\includegraphics[width=8.4 cm]{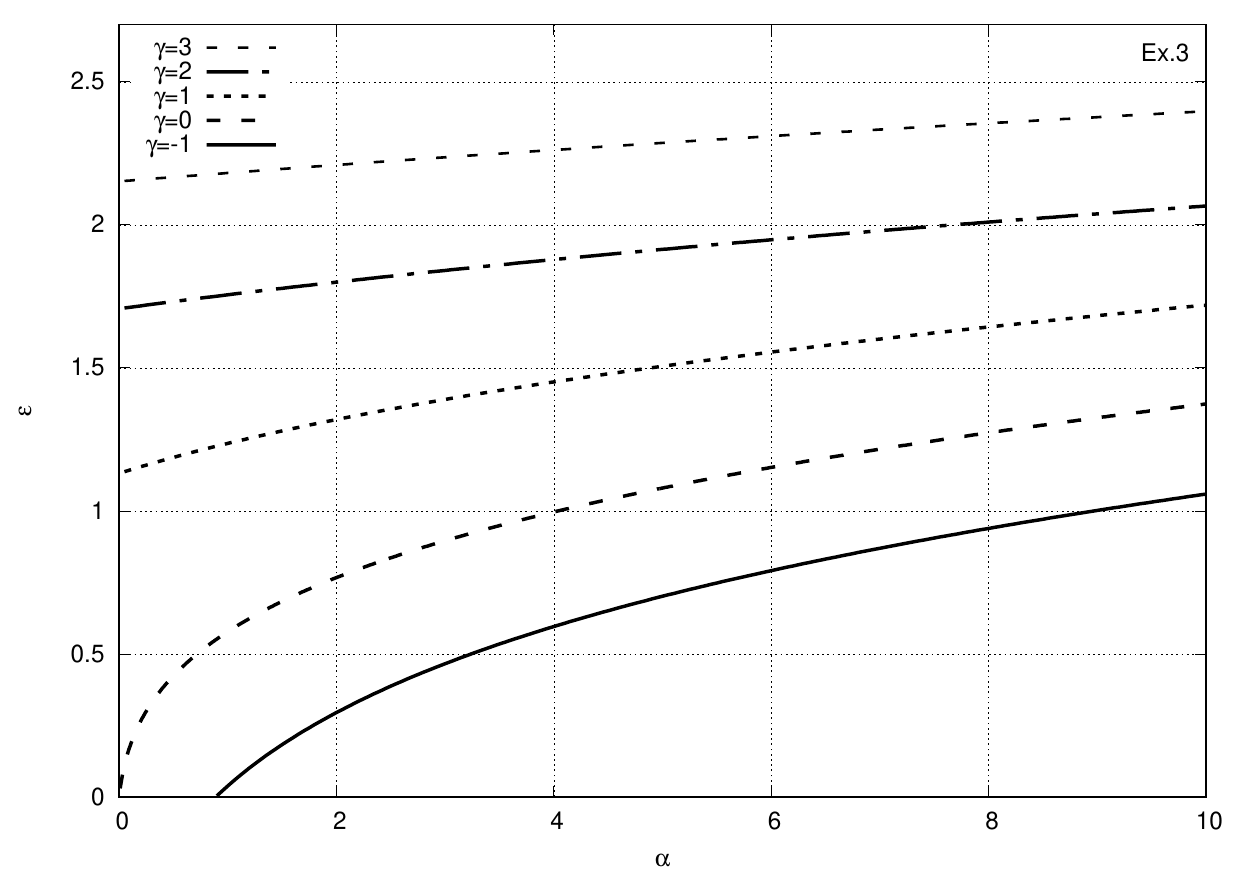}
	\caption{\footnotesize{ Dependencies of the energy level of $\delta-\delta'$ well on coupling constant $\alpha$ for different for different values of $\gamma$ in case of special examples of deformation function} }
	\label{fig2}
\end{figure}
We propose to  write the Schr\"odinger equation for considerable problem as
\be \label{Schroed}
(g(p)^2+q^2)\phi(p)-\frac{\lambda m}{\pi\hbar}\int_{-b}^{b}\phi(p')dp'+\frac{i{\kappa}m}{\pi\hbar^2}\int_{-b}^{b}(p-p')\phi(p')dp'=0.
\ee

Similarly to previous section the solution of the Schr\"odinger equation is assumed to have the form
\be \label{wave}
\psi(p)=\frac{Ap+B}{g^2(p)+q^2}
\ee
Substituting (\ref{wave}) into (\ref{Schroed}) we obtain the following equations
\be
&& A+\frac{i{\kappa}m}{\pi\hbar^2}B\int_{-b}^{b}\frac{dp}{g^2(p)+q^2}=0,\\
&& B-\frac{i{\kappa}m}{\pi\hbar^2}A\int_{-b}^{b}\frac{p^2dp}{g^2(p)+q^2}-\frac{\lambda m}{\pi\hbar}B\int_{-b}^{b}\frac{dp}{g^2(p)+q^2}=0,
\ee
which yields
\be\label{dp_d_energy}
1=\alpha I_1(\epsilon)I_2(\epsilon)+\gamma I_1(\epsilon),
\ee
where $\alpha=\frac{\kappa^2 m^2}{\pi^2\hbar^4}$, $\gamma=\frac{\lambda m}{b\pi\hbar}$  and $I_1(\varepsilon)$ and $I_2(\varepsilon)$ are defined in the previous section.

Remembering that $I_1(\varepsilon)$ and $I_2(\varepsilon)$ are positive and decreasing functions of $\varepsilon$ and $\alpha$ is always positive, we conclude that equation (\ref{dp_d_energy}) has only one solution but in the case of
\be\label{gamma} \gamma>-\gamma_0, \ \  \gamma_0={\alpha}{I_2(0)}>0,\ee with $I_2(0)$ being the maximal value of $I_2(\varepsilon)$.

For the considered in previous section special examples of deformation function  value 
$I_2(0)$ is equal to $2$, $4/3$ and $4\ln2-\pi^2/6\approx1.12765$ correspondingly. From this results we conclude that $\gamma_0$ strongly depends on the choice of the function of deformation. The energy level dependent on $\alpha$ for different values of $\gamma$ is presented on Fig.\ref{fig2}.

\section{Conclusion}
In this paper we have studied the general case of deformed Heisenberg  algebra leading to the minimal length.
The problem of the definition of the $\delta'(x)$ operator has been examined. 
 We  have proposed the definition of $\delta'(x)$ as the lineal kernel of potential energy operator given by (\ref{kernel_dp}). Using this definition  we have solved exactly 1D $\delta'(x)$-potential problem in the general case of deformed Heisenberg algebra leading to the minimal length.  In general case we obtain that energy spectrum consists of only one energy level. In the undeformed limit the  divergency in  the integral $I_1(\varepsilon)$ occurs and energy level
 vanishes. We also have obtained the transcendental equations on the energy spectrum in some  particular cases of the deformation functions. 

We have also considered the case of $-\lambda \delta(x)+\kappa\delta'(x)$ potential. We obtain that there is one bound state for $\kappa>-\kappa_0$ and no bound states for $\kappa\leq-\kappa_0$, with $\kappa_0$ (which is up to notations  given in (\ref{gamma})) depending on $\lambda$ and choice of function of deformation. This fact can serve as a distinguishing factor for different deformation functions.

\section{Acknowledgement}
This work was partly supported by the Project FF-11Hp (No. 0121U100058) from the Ministry of Education and Science of Ukraine and by the Project 2020.02/0196 (No. 0120U104801) from National Research Foundation of Ukraine.

\end{document}